\documentclass[a4paper, preprint]{revtex4}
\usepackage{amsmath}
\usepackage{amsfonts}
\usepackage{graphicx}
\usepackage{color}
\usepackage{lscape}
\usepackage{amssymb}

\begin{document}

\title{Effect of Electron-Electron Interactions on the Charge Carrier Transitions in \textit{trans}-Polyacetylene}
\author{Haibo Ma}
\email{haibo.ma.cn@gmail.com}
\affiliation{Institute of Theoretical and Computational Chemistry,
            Key Laboratory of Mesoscopic Chemistry of MOE,
            School of Chemistry and Chemical Engineering
            Nanjing University,
            Nanjing, 210093,
            China, \\and Institut f\"{u}r
Theoretische Physik C, RWTH Aachen University, D-52056 Aachen,
Germany}
\author{Ulrich Schollw\"{o}ck}
\email{schollwoeck@lmu.de} \affiliation{Department of Physics and Arnold Sommerfeld Center for Theoretical Physics, Ludwig-Maximilians-Universit\"{a}t M\"{u}nchen, D-80333 M\"{u}nchen, Germany, \\and Wissenschaftskolleg zu Berlin, Wallotstrasse 19, D-14193 Berlin, Germany}
\date{Latest revised on \today}

\begin{abstract}

By employing a newly developed nonadiabatic dynamical simulation method, which is a combination of classical molecular
dynamics (MD) and the adaptive time-dependent density matrix
renormalization group (TDDMRG), we investigate the dynamics of charge carrier transitions in \textit{trans}-polyacetylene (PA) with the inclusion of both electron-phonon and electron-electron interactions. The calculations are performed within a modified Su-Schrieffer-Heeger (SSH) model in which electron-electron interactions are taken into account via the combination with extended Hubbard model (EHM). We find that removing an electron from a \textit{trans}-PA chain bearing a positively charged polaron leads to the formation of a pair of charged solitons. Furthermore, we study the effect of electron-electron interactions on such charge carrier transitions in \textit{trans}-PA. Our results show that increasing the on-site Coulomb interaction $U$ and the nearest-neighbor Coulomb repulsion $V$ will not change the qualitative behavior of the transition from a polaron to a soliton pair in the evolution process but will quantitatively reduce the moving velocities of the both formed solitons significantly  and change the conditions for the soliton collisions.

\end{abstract}

\maketitle
\section{Introduction}
Since it was discovered in the 1970s that electrical conductivity
of \textit{trans}-polyacetylene (PA) can be improved significantly
through charge injections or photoexcitations \cite{Chiang77, Shirakawa77, Chiang78}, conjugated polymers have received sustained attentions from both academic and
industrial researchers because of their great numbers of conducting plastic applications, varying from light-emitting diodes, field-effect transistors to photocells and lasers.\cite{MacDiarmid97, Burroughes98, Heeger01, Heeger01_2, Baeriswyl92, Barford05}

Different from the traditional conductors, the charge carriers in conjugated polymers are some self-localized nonlinear excitations, such as solitons, polarons,
or bipolarons, which are inherent features of quasi-one-dimensional conducting
polymer associated with various
structural and charge distortions, depending on the properties of the polymer chain
as well as on the concentration of charge doping.\cite{Heeger01, Heeger01_2, Su79,
Su80, Heeger88}
In \textit{trans}-PA, the most important quasi-particles responsible for the charge transport are solitons serving as domain walls to distinguish the two opposite bond length alternation patterns which might coexist in undisturbed \textit{trans}-PA because of the two-fold ground state energetical degeneracy. A soliton may either be neutral with spin $\pm 1/2$ or present charge $Q=\pm e$ without spin. For the other conjugated polymers without ground state energetical degeneracy, polarons and bipolarons are the basic charge carriers for the charge transport in these materials. A polaron may present charge $Q=\pm e$ with spin $\pm 1/2$ and a bipolaron presents charge $Q=\pm 2e$ with no spin. The studies of the transitions among these different charge carriers are therefore of great interests for the purpose of
modulating electronic preperties of conjugated polymers under different circumstances.

Recently, dynamics of the transitions between the different charge carriers in various conjugated polymers have been theoretically simulated by Silva \textit{et al}.\cite{Lima06, Neto08} It was suggested that in conjugated polymers without ground state energetical degeneracy the most energetically favorable transition is a direct single-polaron to bipolaron transition while in \textit{trans}-PA the most energetically favorable transition is that to a pair of charged solitons. However, in Silva \textit{et al}'s dynamical simulations of charge carrier transitions electron-electron interactions are either completely ignored by only using Su-Schrieffer-Heeger (SSH) model \cite{Su79,
Su80, Heeger88} or merely partially included by adopting the Hartree-Fock (HF) approximation. Actually, many theoretical works have found that electron-electron interactions and electron correlations play a crucial role in determining the nature of electronic excitations in conjugated polymers.\cite{Yonemitsu88, Sim91,
Suhai92, Villar92, Rodriguez-Monge95, Hirata95, Bally92, Fulscher95,
Guo97, Perpete99, Fonseca01, Oliveira03, Champagne04, Monev05, Ma05, Ma05_2,
Ma06, Hu07, Chan05, Chan06, Ghosh08} For example, electron correlation is essential in
determing the correct order of the lowest-lying $\pi\rightarrow\pi^*$ excitations in
\textit{trans}-PA.\cite{Ghosh08} Conjugated polymers behave as quasi-one dimensional systems owing to their strong intramolecular interactions and rather weak intermolecular interactions and electron-electron interactions are accordingly weakly screened. However, accurate simulations of large quantum many-body systems by conventional post-HF methods are still too expensive due to the exponential growth of the number of freedom degrees. Fortunately, recent theoretical studies on the charge carrier transportation dynamics in conjugated polymers by our group and Wu \textit{et al}'s group \cite{Ma08, Ma09, Zhao08, Zhao09} have shown that the combination of classical molecular
dynamics (MD) and the adaptive time-dependent density matrix renormalization group (TDDMRG) \cite{White04, Daley04, Schollwock05} provides a new efficient and accurate approach to perform real time nonadiabatic dynamical simulations of large quantum many-body systems including both electron-lattice and electron-electron interactions.
In this paper, we apply such a method to simulate the charge carrier transition dynamics in \textit{trans}-PA under a photo-ionization of an electron to study the effect of electron-electron interactions on such dynamic processes. During the simulation, the time-independent and time-dependent Schr\"{o}dinger equations are solved within the framework of SSH model modified to include electron-electron interactions via a combination with the extended Hubbard model (EHM).

The paper is organized as follows. In Sec. II, we present the SSH model modified to include electron-electron interactions via EHM and describe the numerical method briefly. In Sec. III the dynamics of the charge carrier transition dynamics in \textit{trans}-PA under a photo-ionization of an electron will be discussed. A summary is given in Sec. IV.

\section{Methodology}
We use the well-known and widely used SSH Hamiltonian \cite{Su79,
Su80, Heeger88} with a combination with the extended Hubbard model (EHM) to describe both the $\pi$-electron part and the lattice backbone with the inclusion of both the electron-lattice interactions and electron-electron interactions:
\begin{equation}\label{H}
    H(t)=H_{el}+H_{latt}.
\end{equation}

The $\pi$-electron part includes both the electron-phonon and the
electron-electron interactions,
\begin{equation}\label{H_el}
\begin{split}
    H_{el}=&-\sum_{<n,n'>,\sigma}t_{n,n'}(c_{n',\sigma}^{+}c_{n,\sigma}+h.c.)\\
& +\frac{U}{2}\sum_{n,\sigma}(c_{n,\sigma}^{+}c_{n,\sigma}-\frac{1}{2})(c_{n,-\sigma}^{+}c_{n,-\sigma}-\frac{1}{2})\\
&
+V\sum_{<n,n'>,\sigma,\sigma'}(c_{n,\sigma}^{+}c_{n,\sigma}-\frac{1}{2})(c_{n',\sigma'}^{+}c_{n',\sigma'}-\frac{1}{2})
\end{split}
\end{equation}
where $t_{n,n'}$ is the hopping integral between the $n$-th site
and the $n'$-th site, while $U$ is the on-site Coulomb interaction
and $V$ denotes the nearest-neighbor electron-electron interaction. $<n,n'>$ denotes the summation is over nearest-neighboring sites.

Because the distortions of the lattice backbone are always within a
certain limited extent, one can adopt a linear relationship between
the hopping integral and the lattice displacements as
$t_{n,n+1}=t_0-\alpha(u_{n+1}-u_n)$ \cite{Su79,
Su80, Heeger88}, where $t_0$
is the hopping integral for zero displacement, $u_n$ the lattice
displacement of the $n$th site, and $\alpha$ is the electron-phonon
coupling.

Because the atoms move much slower than the electrons, we treat the lattice backbone classically with the Hamiltonian
\begin{equation}\label{H_latt}
    H_{latt}=\frac{K}{2}\sum_{n}(u_{n+1}-u_n)^2+\frac{M}{2}\sum_{n}\dot{u}_n^2 \text{    },
\end{equation}
where $K$ is the elastic constant and $M$ is the mass of a site, such as that of a CH monomer for \textit{trans}-PA.

The model parameters are those generally chosen for \textit{trans}-PA:
$t_0$=2.5 eV, $\alpha$=4.1 eV/\AA, $K$=21 eV/\AA$^2$, $M$=1349.14
eVfs$^2$/\AA$^2$, $a$=1.22 \AA, $V=U/2$.\cite{Su79,
Su80, Heeger88, Lima06} We notice that there are some debates on the accuracy of some of the parameters. For example, Ehrenfreund, \textit{et al.} proposed to use $K$=46 eV/\AA$^2$ instead of 21 eV/\AA$^2$ to fit the resonant Raman scattering experiment \cite{Ehrenfreund87}, and an additional linear term was suggested to be included into to Eq.~\ref{H_latt} for the more proper description of the elastic energy of the C-C $\sigma$ backbone \cite{Baeriswyl92} since $a$ is assumed to be the equilibrium lattice spacing of the undimized chain, including both $\sigma$- and $\pi$-bonding. Fur the purpose of making contact to other theoretical studies on charge carrier transition problem
\cite{Lima06, Neto08}, we choose the current parameter schemes. In order to prevent the contraction of the chain, which might result from our simplified lattice elastic energy form, we supplement a constraint of fixed total length of the chain.

The model chain is initially in its ground state configurations for the system containing a charged polaron. Therefore, we need firstly optimize the geometrical configurations of such a singly charged model chain. Due to the exponential growth of the number of degrees of freedom in quantum
many-body systems, the exact solution of our used SSH-EHM model in the complete Hilbert
space is apparently not feasible for our system. Therefore, we perform
approximate solutions by virtue of the density matrix renormalization group (DMRG)
method \cite{White92, White93}. DMRG, firstly proposed by White in 1992 \cite{White92, White93}, which uses the eigenvalues of the subsystem's reduced density matrix as the decimation criterion of Hilbert space, has been shown as an extremely accurate technique in solving one-dimensional strongly correlated system with economic computational costs.\cite{Schollwock05}

Both strong electron-electron and electron-phonon integrations are main features of conducting
polymers and the important reasons for the conjugated polymers to present
novel photoelectronic properties. Therefore, it is obviously not reasonable for the backbone of
conjugated polymers to be frozen in dynamic processes with a time of
more than several femtoseconds. On the other way, if all the carbon-carbon
bond lengths are fixed in a undimerized way, no solitons will be
formed because there are no degenerate ground state possibilities. There
might be also some charge density waves in such conditions, however, they
will move forth and back much more quickly than the normal solitons and of course they are not the real case in conjugated polymers.

For the purpose of performing real-time simulation of both the evolution of quantum $\pi$-electron part and the classical movement of the chain backbone, we adopt a newly developed real-time simulation method in which classical molecular dynamics is combined with the adaptive TDDMRG \cite{White04, Daley04, Schollwock05}. The main idea of this method is to evolve the $\pi$-electron part by the adaptive TDDMRG and move the backbone part by classical MD iteratively and nonadiabatically to include nearly all relevant electron-phonon and
electron-electron interactions. Details about such the combined TDDMRG/MD method can be found in the recent papers \cite{Ma08, Ma09, Zhao08, Zhao09}.

\section{Results and discussion}
The simulated system consists of a \textit{trans}-PA chain with 100 (CH) monomers, where a positively charged polaron is initially localized in the middle of the chain. After the time begins to evolve, one more electron is removed through photoionization. Then the dynamics of the charge carrier transitions in \textit{trans}-PA within 400 femtoseconds (fs) is simulated by virtue of classical MD combined with the adaptive TDDMRG.

Recent works have carefully studied the accuracy performance of the combined TDDMRG/MD method and verified its reliability with suitable adopted parameters and it was found that one can achieve reliable and converged numerical results if the DMRG truncation weight $\epsilon_{\rho}$ is taken small enough
($\epsilon_{\rho}\leq1.0\times10^{-6}$) and suitable time step $\Delta t$ values around 0.01$\sim$0.05 fs are chosen.\cite{Ma08, Ma09, Zhao08, Zhao09}  All our numerical results presented in this work are well-converged values calculated with $\epsilon_{\rho}=1.0\times10^{-7}$ and $\Delta t$=0.05 fs.

\subsection{Structure of the initial state}
At $t=0$, the singly charged model \textit{trans}-PA chain with 100 (CH) monomers is located at its ground state. Its optimized geometric structure and charge distribution pattern at $U$=2.0 eV and $V$=1.0 eV are shown in Fig.~\ref{fig:ge0} and
Fig.~\ref{fig:cd0}.
Considering there are strong bond length alternations and accordingly strong charge density oscillations due to significant electron-phonon and electron-electron interactions, for the purpose of presenting a better visualization of the results with more smooth curves, we use a mean charge density $\bar{\rho}_n(t)$, derived from the charge density $\rho_n(t)$, and the staggered bond order
parameter $r_n(t)$ by taking the average values of nearest-neighboring sites to analyze the simulations \cite{Rakhmanova99},
\begin{equation}\label{rho}
    \bar{\rho}_n(t)=\frac{\rho_{n-1}(t)+2\rho_n(t)+\rho_{n+1}(t)}{4},
\end{equation}
\begin{equation}\label{bo}
    r_n(t)=(-1)^n\frac{u_{n-1}(t)+2u_n(t)+u_{n+1}(t)}{4},
\end{equation}
where $\rho_n(t)=1-\sum_{\sigma}\langle\Psi(t)|(c_{n,\sigma}^{+}c_{n,\sigma}|\Psi(t)\rangle$ and $u_n$ is the lattice
displacement of the $n$th site as mentioned earlier.
One may find typical characteristics of polarons in conducting polymers \cite{Heeger88} from the geometric structure and charge distribution pattern before photoionization as illustrated in Fig.~\ref{fig:ge0} and
Fig.~\ref{fig:cd0}. This implies that the initial singly charged model chain is bearing a polaron defect.
The model chain is then assumed to undergo further vertical photoionization in the spirit of the Franck-Condon principle, which means one more electron is taken away from the original singly charged model system immediately at the unchanged ground state geometry of singly charged model chain.
The removal of the second electron originally located at singly occupied molecular orbital (SOMO) makes the new doubly ionized state lying at higher energy level than the ground state of the original singly charged system. Such kind of energy difference is determined to be about 0.84 eV by our calculations with $U$=2.0 eV and $V$=1.0 eV, implying that such doubly ionized state locates between the second and third lowest doublet excited states of singly charged \textit{trans}-PA, whose vertical excitation energies were predicted to be about 0.58 and 0.91 eV respectively with Pariser-Parr-Pople (PPP) calculations \cite{Ma05_2}.

\begin{center}
(Fig.~\ref{fig:ge0} about here)
\end{center}

\begin{center}
(Fig.~\ref{fig:cd0} about here)
\end{center}

\subsection{General picture of charge carrier transitions in \textit{trans}-PA}
\begin{center}
(Fig.~\ref{fig:cd_2_1} about here)
\end{center}

\begin{center}
(Fig.~\ref{fig:ge_2_1} about here)
\end{center}

Now let us show the general time evolution
pictures of charge density and site displacement in the charge carrier transition process in \textit{trans}-PA.
The time evolution pictures of the mean charge density $\bar{\rho}_n(t)$ and the staggered bond order
parameter $r_n(t)$ are shown in Fig.~\ref{fig:cd_2_1} and
Fig.~\ref{fig:ge_2_1} respectively.

As we can see from Fig.~\ref{fig:cd_2_1}, the removal of an electron from the \textit{trans}-PA chain which initially bears a positively charged polaron induces larger charge densities in the central area of the chain. Then the induced large charged densities in the middle chain tend to disperse to the nearby areas toward the both ends of the chain while the geometrical distortions will also evolve as illustrated in Fig.~\ref{fig:ge_2_1} due to the electron-phonon interactions. Because the two opposite bond length alternation patterns are energetically degenerate in \textit{trans}-PA, it is energetically more favorable for the new excitation to break into a pair of charged solitons than a pair of polarons. It could be found that a pair of charged solitons with two charge peaks have been formed after about 20 fs from the pictures of the charge density and geometrical structure in Fig.~\ref{fig:cd_2_1} and
Fig.~\ref{fig:ge_2_1}. Clearly, the two charged solitons repel each other because they are with the same charge sign.
From Fig.~\ref{fig:cd_2_1} we may find that the moving charge peaks do not really change at all
(but for the positions) during the repulsion processes of the two new defects which shows that the transition from polaron to charged soliton pair has been finished and the two newly formed charged solitons can move relatively freely after 20 fs. From
Fig.~\ref{fig:ge_2_1} it is found that during the movement of the two charged solitons the center positions of the solitons move as nearly strait lines with time evolving which implies the solitons are moving with a nearly constant velocity. It could also be found that the charge distortion is always coupled with the geometrical distortions very well during the entire moving process of the soliton which verifies that the soliton structure is an inherent feature of \textit{trans}-PA. In Fig.~\ref{fig:ge_2_1} we can notice some wave grains appear after the solitons begin moving. That happens because there is always a long-lasting
oscillatory ``tail'' appearing behind the soliton defect center during the soliton moving process.\cite{Ma08} This
``tail'' is generated by the inertia of those monomers to fulfill
energy and momentum conservation and it absorbs the additional
energy, preventing the further increasement of the soliton velocity
after a stationary value is reached.

As illustrated in Fig.~\ref{fig:cd_2_1} and
Fig.~\ref{fig:ge_2_1}, the two charged solitons seem to be able to move along the chain back and forth independently after they are formed at around 20 fs. They get reflections when they encounter the boundaries of the chain at around 140 fs, and then they return back and get the collisions when they meet each other at around 280 fs. Afterwards, the charged soliton pair will repeat the processes of repulsion, independent moving and collision.

\subsection{Effect of electron-electron interactions on the charge carrier transitions in \textit{trans}-PA}

In order to study the influence of electron-electron interactions
on the charge carrier transitions in \textit{trans}-PA, we perform various simulations for the charge carrier transition dynamics with different adopted values for the electron-electron interaction strengths. Time evolution pictures of the mean charge density $\bar{\rho}_n(t)$ and the staggered bond order
parameter $r_n(t)$ simulated with different the on-site Coulomb interaction $U$ values and the nearest-neighbor Coulomb repulsion $V$ values are displayed in Fig.~\ref{fig:cd} and Fig.~\ref{fig:ge}. 

\begin{center}
(Fig.~\ref{fig:cd} about here)
\end{center}

\begin{center}
(Fig.~\ref{fig:ge} about here)
\end{center}

From Fig.~\ref{fig:cd} and Fig.~\ref{fig:ge} we can clearly see that the evolution pictures for both the charge densities and the geometrical parameters with different electron-electron interaction strengths are qualitatively very similar. This means the varying of $U$ and $V$ will not change the qualitative behavior of the transition from a polaron to a soliton pair in \textit{trans}-PA after photo-ionization: removing an electron from the chain initially bearing a positively charged polaron leads to the formation of a pair of charged solitons and then the formed two solitons will move back and forth along the chain independently as well as two solitons will get reflections at the chain boundaries and get collisions when encountering each other in the chain center region.

However, we may found that increasing both $U$ and $V$ will quantitatively reduce the moving velocities of the two newly formed solitons as illustrated in Fig.~\ref{fig:cd} and Fig.~\ref{fig:ge}. Such an observation is in good agreement with our previous studies on the soliton transport dynamics in \textit{trans}-PA under an external electric field \cite{Ma08}, in which we have found that in most cases ($U\geq0.9$ eV) increasing $U$ will induce the reduction of electron delocalization and accordingly it will suppress the transport of the charged soliton. Our previous work has also shown that increasing $U$ when $U\leq0.9$ eV will favor the charged soliton transport while increasing $V$ when $V$ is small will be very unbeneficial to the charged soliton transport because it will induce a
more localized defect distribution.\cite{Ma08} Therefore, it is reasonable to see increasing both $U$ and $V$ at the same time will suppress the movement of two newly formed charged solitons, as illustrated in Fig.~\ref{fig:cd} and Fig.~\ref{fig:ge}.

At the same time, we may also notice the varying of $U$ and $V$ will change the conditions for the collision of the two newly formed charged solitons from Fig.~\ref{fig:cd} and Fig.~\ref{fig:ge}. With the enhanced $U$ and $V$, the minimal distance between the two charged solitons during the collision process will be significantly increased, from less than 10 units in the case with $U$=0.0 eV and $V$=0.0 eV to about 30 units in that with $U$=3.0 eV and $V$=1.5 eV. This can be easily understood for the fact that the enhanced electron-electron repulsions will make the repulsion between two charged solitons stronger and accordingly the two charged solitons can hardly move furthermore to touch each other completely under the case with large $U$ and $V$ values.

Actually, the case with $U$=0 and $V$=0 in our presented results is the same as what has been studied in Silva \emph{et al}¡¯ work with pure SSH model\cite{Neto08}. As we have disclosed before, the further consideration of the on-site Couloumb repulsions $U$ and the nearest-neighbor electron-electron interactions $V$ will not change the qualitative behavior of the charge carrier transition in \textit{trans}-PA after photo-ionization, but will reduce the moving velocities of the two newly formed solitons significantly as a result of decreasing electron delocalization.

In this work, we use $V$ values varying from 0 to only 1.5 eV which might be not so realistic in view of the spectroscopic properties of ethylene. The further increase of nearest-neighbor
interactions $V$ will strengthen the electron-hole attractions and accordingly
induce much more delocalized soliton picture according to previous studies\cite{Ma08}. Considering our current
model chain has only 100 sites and the increase of $U$ and $V$ do not change the qualitative behavior of the charge carrier transition in \textit{trans}-PA at least within this study, we don't use larger $U$ and $V$ values in this work for the fact that it is difficult to present the pictures of two moving and much more delocalized solitons very clearly in a short chain. However, we are expecting to simulate larger model systems with more realistic
electronic integrals with more powerful computational facilities in the future works.

\section{Summary}

For a model \textit{trans}-PA chain initially holding a polaron defect, described with a modified SSH model in which electron-electron interactions are included via the combination with EHM, we
have studied the dynamics of charge carrier transitions in such a \textit{trans}-PA chain after photo-ionization. Nearly all relevant electron-phonon and electron-electron interactions have been
fully taken into account by solving the time-dependent Schr\"{o}dinger equation for the $\pi$-electrons and the Newton's equation of motion for the backbone monomer displacements with the combination of the adaptive TDDMRG and classical MD.

It is found that removing an electron from the \textit{trans}-PA chain bearing a positively charged polaron leads to the formation of a pair of charged solitons and then the formed two solitons will move back and forth along the chain independently. These two solitons will get reflections at the chain boundaries and get collisions when encountering each other in the chain center region.

Furthermore, we study the influence of electron-electron interactions on such charge carrier transitions in \textit{trans}-PA. Our results show that increasing the on-site Coulomb interaction $U$ and the nearest-neighbor Coulomb repulsion $V$ will not change the qualitative behavior of the transition from a polaron to a soliton pair in the evolution process. However, increasing both $U$ and $V$ will quantitatively reduce the moving velocities of the two newly formed solitons significantly as a result of decreasing electron delocalization and increasing $U$ and $V$ will also change the conditions for the soliton collisions.

\section*{Acknowledgment}
HM acknowledges the support by Alexander von
Humboldt Research Fellowship.

\newpage

\begin{figure}
\caption{\label{fig:ge0} Ma \textit{et al}, J. Phys. Chem.}
\includegraphics[width =8 cm]{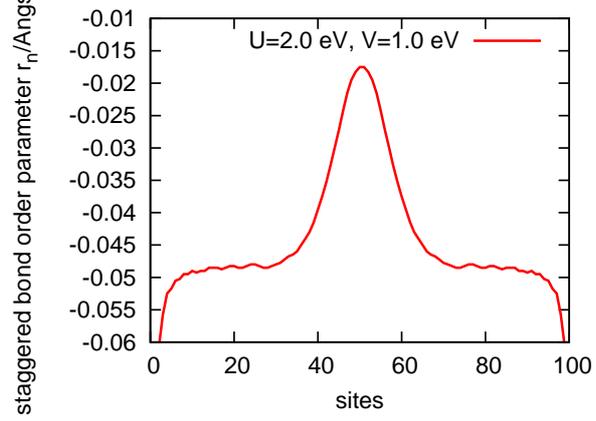}
\end{figure}

\begin{figure}
\caption{\label{fig:cd0} Ma \textit{et al}, J. Phys. Chem.}
\includegraphics[width =8 cm]{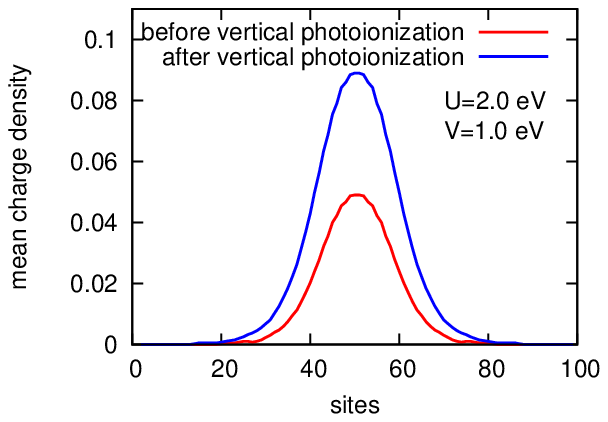}
\end{figure}

\begin{figure}
\caption{\label{fig:cd_2_1} Ma \textit{et al}, J. Phys. Chem.}
\includegraphics[width =8 cm]{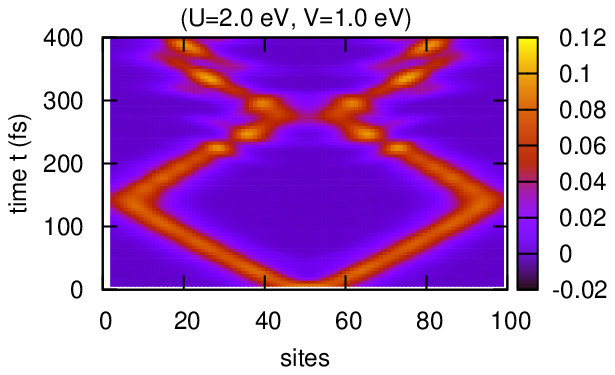}
\end{figure}

\begin{figure}
\caption{\label{fig:ge_2_1} Ma \textit{et al}, J. Phys. Chem.}
\includegraphics[width =8 cm]{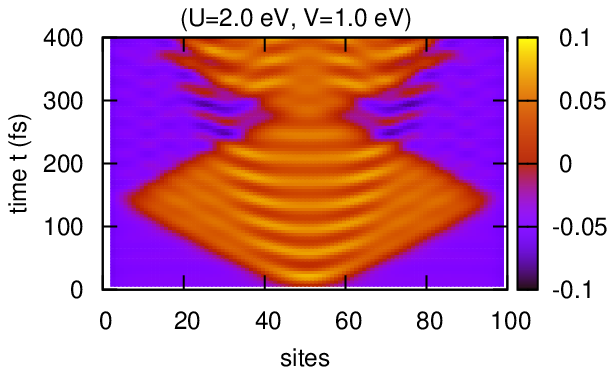}
\end{figure}

\begin{figure}
    \caption{\label{fig:cd} Ma \textit{et al}, J. Phys. Chem.}
  \begin{center}
    \begin{tabular}{cc}
      \resizebox{80mm}{!}{\includegraphics{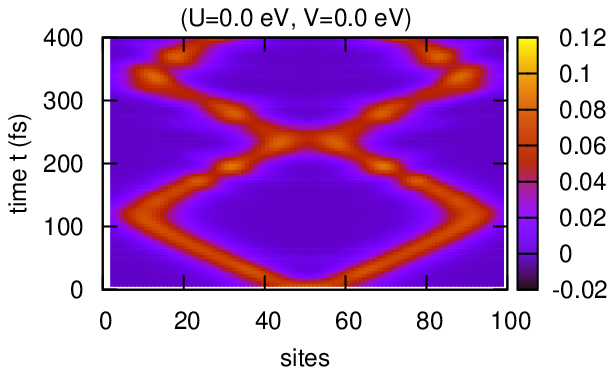}} &
      \resizebox{80mm}{!}{\includegraphics{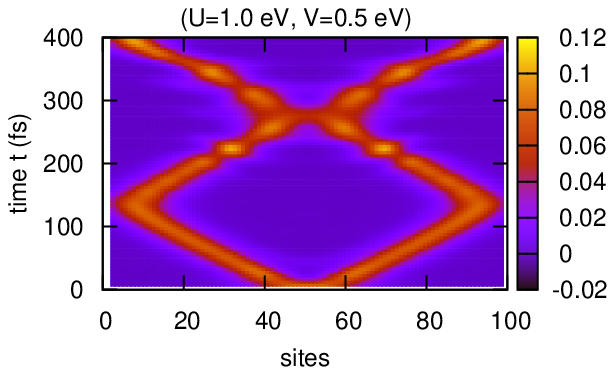}} \\
      \resizebox{80mm}{!}{\includegraphics{cd_2.eps}} &
      \resizebox{80mm}{!}{\includegraphics{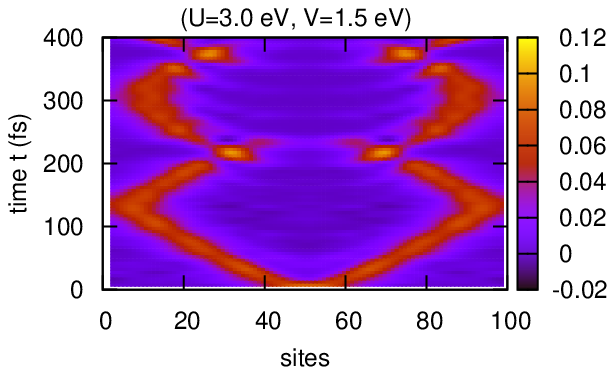}} \\
    \end{tabular}
  \end{center}
\end{figure}

\begin{figure}
    \caption{\label{fig:ge} Ma \textit{et al}, J. Phys. Chem.}
  \begin{center}
    \begin{tabular}{cc}
      \resizebox{80mm}{!}{\includegraphics{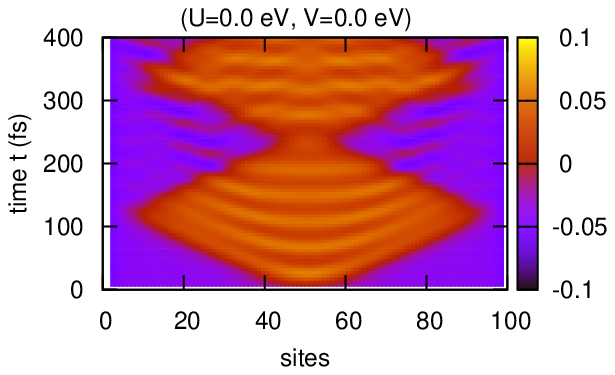}} &
      \resizebox{80mm}{!}{\includegraphics{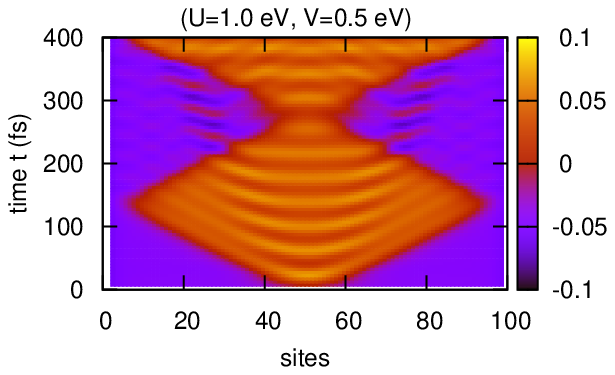}} \\
      \resizebox{80mm}{!}{\includegraphics{ge_2.eps}} &
      \resizebox{80mm}{!}{\includegraphics{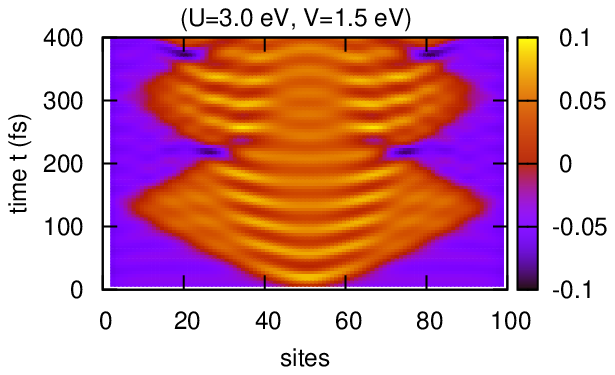}} \\
    \end{tabular}
  \end{center}
\end{figure}

\noindent \textbf{Figure Captions}
\newline
\newline
FIG.~\ref{fig:ge0} Geometric structure at the initial state. ($U$=2.0 eV, $V$=1.0 eV)
\newline
\newline
FIG.~\ref{fig:cd0} Charge distribution picture before and after vertical photoionization at the initial state. ($U$=2.0 eV, $V$=1.0 eV)
\newline
\newline
FIG.~\ref{fig:cd_2_1} Time evolution of the mean charge density $\bar{\rho}_n(t)$. ($U$=2.0 eV, $V$=1.0 eV)
\newline
\newline
FIG.~\ref{fig:ge_2_1} Time evolution of the staggered bond order
parameter $r_n(t)$. ($U$=2.0 eV, $V$=1.0 eV)
\newline
\newline
FIG.~\ref{fig:cd} Time evolution of the mean charge density $\bar{\rho}_n(t)$ with various electron-electron interaction parameters.
\newline
\newline
FIG.~\ref{fig:ge} Time evolution of the staggered bond order
parameter $r_n(t)$ with various electron-electron interaction parameters.
\newline

\clearpage

\clearpage

\clearpage

\clearpage

\clearpage

\clearpage


\section*{References}
\begin{thebibliography}{}

\bibitem{Chiang77} 
C. K. Chiang, C. R. Fincher, Jr., Y. W. Park, A. J. Heeger, H. Shirakawa, E. J. Louis, S. C. Gau, and Alan G. MacDiarmid,
Phys. Rev. Lett. \textbf{39}, 1098 (1977).

\bibitem{Shirakawa77}
H. Shirakawa, E. J. Louis, Alan G. MacDiarmid, C. K. Chiang, and A. J. Heeger, J. Chem. Soc., Chem. Commun.
\textbf{16}, 579 (1977).

\bibitem{Chiang78} 
C. K. Chiang, M. A. Druy, S. C. Gau, A. J. Heeger, E. J. Louis, Alan G. MacDiarmid, Y. W. Park, and H. Shirakawa, J. Am. Chem.
Soc. \textbf{100}, 1013 (1978).

\bibitem{MacDiarmid97}
A. G. MacDiarmid, Synth. Met. \textbf{84}, 27 (1997).

\bibitem{Burroughes98}
J. H. Burroughes, C. A. Jones, and R. H. Friend, Nature \textbf{335}, 137 (1998).

\bibitem{Baeriswyl92}
D. Baeriswyl, D. K. Campbell, and S. Mazumdar, \textit{Conjugated Conducting Polymers (H. Kiess. ed.)} (Springer-Verlag, Berlin, 1992).

\bibitem{Barford05}
W. Barford, \textit{Electronic and Optical Properties of Conjugated Polymers} (Oxford University Press, Oxford, 2005).

\bibitem{Heeger01}
A. J. Heeger, J. Phys. Chem. B \textbf{105}, 18475 (2001).

\bibitem{Heeger01_2}
A. J. Heeger, Rev. Mod. Phys. \textbf{73}, 681 (2001).

\bibitem{Su79} 
W. P. Su, J. R. Schrieffer, and A. J. Heeger, Phys. Rev.
Lett. \textbf{42}, 1698 (1979).

\bibitem{Su80}
W. P. Su, J. R. Schrieffer, and A. J. Heeger, Phys. Rev. B \textbf{22}, 2099 (1980).

\bibitem{Heeger88}
A. J. Heeger, S. Kivelson, J. R. Schrieffer, and W. P. Su, Rev. Mod. Phys. \textbf{60}, 781 (1988).

\bibitem{Lima06}
M. P. Lima and G. M. E. Silva, Phys. Rev. B \textbf{74},
224304 (2006).

\bibitem{Neto08}
P. H. D. Neto, W. F. da Cunha, R. Gargano and G. M. E. Silva,  Int. J. Quantum. Chem. \textbf{108},
2507 (2008).


\bibitem{Yonemitsu88} 
K. Yonemitsu, Y. Ono, and Y. Wada, J. Phys. Soc. Jpn. \textbf{57}, 3875 (1988).

\bibitem{Sim91} 
F. Sim, D. R. Salahub, S. Chin, and M. Dupuis, J. Chem.
Phys. \textbf{95}, 4317 (1991).

\bibitem{Suhai92} 
S. Suhai, Int. J. Quantum. Chem.
\textbf{42}, 193 (1992).

\bibitem{Villar92} 
H. O. Villar and M. Dupuis, Theor. Chim. Acta. \textbf{83}, 155 (1992).

\bibitem{Bally92} 
T. Bally, K. Roth, W. Tang, R. R. Schrock, K. Knoll, and L.
Y. Park, J. Am. Chem. Soc. \textbf{114}, 2440 (1992).

\bibitem{Rodriguez-Monge95} 
L. Rodr\'{i}guez-Monge and S. Larsson, J. Chem. Phys.
\textbf{102}, 7106 (1995).

\bibitem{Hirata95}
S. Hirata, H. Torii, and M. Tasumi, J. Chem. Phys. \textbf{103}, 8964 (1995).

\bibitem{Fulscher95} 
M. P. F\"{u}lscher, S. Matzinger, and T. Bally, Chem.
Phys. Lett. \textbf{236}, 167 (1995).

\bibitem{Guo97}
H. Guo and J. Paldus, Int. J. Quantum. Chem. \textbf{63}, 345 (1997).

\bibitem{Perpete99}
E. A. Perp\`{e}te and B. Champagne, J. Mol. Struct.
(THEOCHEM) \textbf{487}, 39 (1999).

\bibitem{Fonseca01} 
T. L. Fonseca, M. A. Castro, C. Cunha, and O. A. V. Amaral,
Synth. Met. \textbf{123}, 11 (2001).

\bibitem{Oliveira03} 
L. N. Oliveira, O. A. V. Amaral, M. A. Castro, and T. L. Fonseca,
Chem. Phys. \textbf{289}, 221 (2003).

\bibitem{Champagne04} 
B. Champagne and M. Spassova, Phys. Chem. Chem. Phys.
\textbf{6}, 3167 (2004).

\bibitem{Monev05}
V. Monev, M. Spassova, and B. Champagne, Int. J. Quantum.
Chem. \textbf{104}, 354 (2005).

\bibitem{Ma05}
H. Ma, F. Cai, C. Liu, and Y. Jiang, J. Chem. Phys.
\textbf{122}, 104909 (2005).

\bibitem{Ma05_2}
H. Ma, C. Liu, and Y. Jiang, J. Chem. Phys.
\textbf{123}, 084303 (2005).

\bibitem{Ma06}
H. Ma, C. Liu, and Y. Jiang, J. Phys. Chem. B
\textbf{110}, 26488 (2006).

\bibitem{Hu07}
W. Hu, H. Ma, C. Liu, and Y. Jiang, J. Chem. Phys.
\textbf{126}, 044903 (2007).

\bibitem{Chan05}
G. K. L. Chan and T. Van Voorhis, J. Chem. Phys.
\textbf{122}, 204101 (2005).

\bibitem{Chan06}
J. Hachmann, W. Cardoen and  G. K. L. Chan, J. Chem. Phys.
\textbf{125}, 144101 (2006).

\bibitem{Ghosh08}
D. Ghosh, J. Hachmann, T. Yanai and G. K. L. Chan, J. Chem. Phys.
\textbf{128}, 144117 (2008).

\bibitem{Ma08}
H. Ma and U. Schollw\"{o}ck, J. Chem. Phys. \textbf{129}, 244705 (2008).

\bibitem{Ma09}
H. Ma and U. Schollw\"{o}ck, J. Phys. Chem. A  \textbf{113}, 1360 (2009).

\bibitem{Zhao08}
H. Zhao, Y. Yao, Z. An and C. Wu, Phys. Rev. B \textbf{78}, 035209 (2008).

\bibitem{Zhao09}
H. Zhao, Y. Chen, X, Zhang, Z. An and C. Wu, J. Chem. Phys. \textbf{130}, 234908 (2009).


\bibitem{White04}
S. R. White and A. Feiguin, Phys. Rev. Lett. \textbf{93}, 076401 (2004).

\bibitem{Daley04}
A. J. Daley, C. Kollath, U. Schollw\"{o}ck, and G. Vidal, J. Stat. Mech. P04005 (2004).

\bibitem{Schollwock05}
U. Schollw\"{o}ck, Rev. Mod. Phys. \textbf{77}, 259 (2005).

\bibitem{Ehrenfreund87}
E. Ehrenfreund, Z. Vardeny, O. Brafman, and B. Horovitz, Phys. Rev. B
\textbf{36}, 1535 (1987).

\bibitem{White92}
S. R. White, Phys. Rev. Lett. \textbf{69},
2863 (1992).

\bibitem{White93}
S. R. White, Phys. Rev. B \textbf{48},
10345 (1993).

\bibitem{Rakhmanova99}
S. V. Rakhmanova, and E. M. Conwell, Appl. Phys. Lett. \textbf{75},
1518 (1999).

\end{thebibliography}
\end{document}